\documentstyle[tighten,preprint,pra,aps]{revtex}
\begin{document}
\draft
\title{Comment on ``Accurate relativistic effective potentials
       for the sixth-row main group elements''
       \protect[J.\ Chem.\ Phys.\ 107, 9975 (1997)\protect].}
\author{N.~S.~Mosyagin~\cite{E-mail} and A.~V.~Titov}
\address{Petersburg Nuclear Physics Institute, \\
         Gatchina, Petersburg district 188350, Russia}
\date{\today}
\maketitle

\begin{abstract}
Critical remarks in respect to the Generalized Relativistic
Effective Core Potential in the recent article of S.~A.~Wildman at al.\
are discussed and shown to be incorrect.
\end{abstract}

\vspace{1cm}

S.~A.~Wildman, G.~A.~DiLabio, and P.~A.~Christiansen's article~\cite{Wildman}
deals with a problem of the Relativistic Effective Core Potential (RECP)
generation for precise calculations. In this article, new RECPs for the Tl,
Pb, Bi, At, and Rn atoms which are more accurate than the
previous RECPs~\cite{Ross} are constructed and tested in molecular
calculations. However, some statements made in the article are
incorrect.

As it follows from article~\cite{Wildman}, the problems with the old
RECPs~\cite{Ross} ``$\ldots$~are due to the inappropriate partitioning of
the spinors used to generate the $f_{5/2}$ and $f_{7/2}$ potentials''
that, in its turn, is connected with the restriction on the number ($n\leq 2$)
of inflections for pseudospinors in the RECP generation procedure
developed in~\cite{Christ}.  However, this does not mean that
``Given the comparable approaches in terms of orbital/spinor partitioning,
one would expect to see very similar behavior and problems with
the potentials developed by other groups$^{16}$~$\ldots$'' (i.e.,
Ref.~\cite{Tupitsyn} here).  Generalized RECP (GRECP) generation
procedure~\cite{Tupitsyn} is based on procedure~\cite{Christ} but has
essential distinctions. Some of them are requirement of a smooth shape for
the potential and absence of limitations on the number of inflections for
pseudospinors~\cite{Tupitsyn}. We did not concentrate on such technical
details in our papers, because they are not so important as the principal
questions concerning the optimal form of the RECP operator.  One can see
from comparison of Fig.~\ref{Pb_5f} in this comment with Fig.~1(B) in
Ref.~\cite{Wildman}, however, that the
new $5f_{5/2}$ pseudospinor of S.~A.~Wildman et al., in contrast to the
old one, is rather close to the $5f_{5/2}$ pseudospinor constructed within
the GRECP generation procedure~\cite{Tupitsyn}.

Some comments should be made on the authors' statement that in
paper~\cite{Tupitsyn} ``$\ldots$~it has occasionally been suggested that
the small components are neglected in this procedure~$\ldots$'' (i.e.\ in
the GRECP generation procedure). More correctly (that is evident from our
papers, see e.g.~\cite{Titov1}), we neglect contribution from the
small components in the valence region to the two-electron integrals and
the electronic density, and only the large components of Dirac spinors are
used in order to construct the pseudospinors. In paper~\cite{Tupitsyn}, it
was also specified that we applied the two-component technique of Lee et
al.~\cite{Lee} where the authors, in particular, wrote at the end of page
5862: ``Since small components become even less important in the outer
region, it may be reasonable to assume that {\bf small components can be
neglected} in calculations that emphasize the description of valence
electrons''. Nevertheless, it is clear from this paper that the authors
understand that the
small components can not be neglected in Dirac-like equations.  A few
lines above in the commented article, S.~A.~Wildman et al.\
write: ``The two-component
pseudospinor, $\chi_{lj}$, is then effectively reinserted into the radial
DHF equation and the equation is inverted to obtain the localized
relativistic effective potential, $U^{REP}_{lj}(r)$''. In reality, one
should use ``nonrelativistic-type'' HF equations (i.e.\ containing the
nonrelativistic kinetic energy operator) in the $jj$-coupling scheme which
are augmented with the relativistic $j$-dependent potentials as it is made
in~\cite{Lee}.

 The work on development of the GRECP method was supported by the DFG/RFBR
 grant N 96--03--00069, the RFBR grant N 96--03--33036 and by the INTAS grant
 No 96--1266.

\unitlength=0.78pt
\begin{figure}
\begin{center}
\vspace{6cm}
\begin{picture}(435,325)( -29.977, -62.805)
\put(   0.023, -42.805){\vector(0,1){300}}
\put(   5.023, 257.195){\makebox(0,0)[tl]{$\chi_{5f_{5/2}}(r)\times r$,
                                          $\phi_{5f_{5/2}}(r)\times r$ (a.u.)}}
\put(   0.023, -42.805){\vector(1,0){400}}
\put( 400.023, -37.805){\makebox(0,0)[br]{$r$ (a.u.)}}
%\put(   0.023, -42.805){\line(0,-1){3}}
%\put(   0.023, -47.805){\makebox(0,0)[t]{ 0.00}}
%\put(  40.023, -42.805){\line(0,-1){3}}
%\put(  40.023, -47.805){\makebox(0,0)[t]{ 0.40}}
%\put(  80.023, -42.805){\line(0,-1){3}}
%\put(  80.023, -47.805){\makebox(0,0)[t]{ 0.79}}
%\put( 120.023, -42.805){\line(0,-1){3}}
%\put( 120.023, -47.805){\makebox(0,0)[t]{ 1.19}}
%\put( 160.023, -42.805){\line(0,-1){3}}
%\put( 160.023, -47.805){\makebox(0,0)[t]{ 1.59}}
%\put( 200.023, -42.805){\line(0,-1){3}}
%\put( 200.023, -47.805){\makebox(0,0)[t]{ 1.98}}
%\put( 240.023, -42.805){\line(0,-1){3}}
%\put( 240.023, -47.805){\makebox(0,0)[t]{ 2.38}}
%\put( 280.023, -42.805){\line(0,-1){3}}
%\put( 280.023, -47.805){\makebox(0,0)[t]{ 2.78}}
%\put( 320.023, -42.805){\line(0,-1){3}}
%\put( 320.023, -47.805){\makebox(0,0)[t]{ 3.17}}
%\put( 360.023, -42.805){\line(0,-1){3}}
%\put( 360.023, -47.805){\makebox(0,0)[t]{ 3.57}}
\put(   0.023, -42.805){\line(0,-1){3}}
\put(   0.023, -47.805){\makebox(0,0)[t]{ 0.0}}
\put(  50.443, -42.805){\line(0,-1){3}}
\put(  50.443, -47.805){\makebox(0,0)[t]{ 0.5}}
\put( 100.863, -42.805){\line(0,-1){3}}
\put( 100.863, -47.805){\makebox(0,0)[t]{ 1.0}}
\put( 151.284, -42.805){\line(0,-1){3}}
\put( 151.284, -47.805){\makebox(0,0)[t]{ 1.5}}
\put( 201.704, -42.805){\line(0,-1){3}}
\put( 201.704, -47.805){\makebox(0,0)[t]{ 2.0}}
\put( 252.124, -42.805){\line(0,-1){3}}
\put( 252.124, -47.805){\makebox(0,0)[t]{ 2.5}}
\put( 302.544, -42.805){\line(0,-1){3}}
\put( 302.544, -47.805){\makebox(0,0)[t]{ 3.0}}
\put( 352.964, -42.805){\line(0,-1){3}}
\put( 352.964, -47.805){\makebox(0,0)[t]{ 3.5}}
%\put(   0.023, -42.805){\line(-1,0){3}}
%\put(  -4.977, -42.805){\makebox(0,0)[r]{-.050}}
%\put(   0.023, -12.805){\line(-1,0){3}}
%\put(  -4.977, -12.805){\makebox(0,0)[r]{-.015}}
%\put(   0.023,  17.195){\line(-1,0){3}}
%\put(  -4.977,  17.195){\makebox(0,0)[r]{ .020}}
%\put(   0.023,  47.195){\line(-1,0){3}}
%\put(  -4.977,  47.195){\makebox(0,0)[r]{ .055}}
%\put(   0.023,  77.195){\line(-1,0){3}}
%\put(  -4.977,  77.195){\makebox(0,0)[r]{ .090}}
%\put(   0.023, 107.195){\line(-1,0){3}}
%\put(  -4.977, 107.195){\makebox(0,0)[r]{ .125}}
%\put(   0.023, 137.195){\line(-1,0){3}}
%\put(  -4.977, 137.195){\makebox(0,0)[r]{ .160}}
%\put(   0.023, 167.195){\line(-1,0){3}}
%\put(  -4.977, 167.195){\makebox(0,0)[r]{ .195}}
%\put(   0.023, 197.195){\line(-1,0){3}}
%\put(  -4.977, 197.195){\makebox(0,0)[r]{ .229}}
%\put(   0.023, 227.195){\line(-1,0){3}}
%\put(  -4.977, 227.195){\makebox(0,0)[r]{ .264}}
\put(   0.023, -42.805){\line(-1,0){3}}
\put(  -4.977, -42.805){\makebox(0,0)[r]{-.05}}
\put(   0.023,   0.189){\line(-1,0){3}}
\put(  -4.977,   0.189){\makebox(0,0)[r]{ .00}}
\put(   0.023,  43.182){\line(-1,0){3}}
\put(  -4.977,  43.182){\makebox(0,0)[r]{ .05}}
\put(   0.023,  86.176){\line(-1,0){3}}
\put(  -4.977,  86.176){\makebox(0,0)[r]{ .10}}
\put(   0.023, 129.170){\line(-1,0){3}}
\put(  -4.977, 129.170){\makebox(0,0)[r]{ .15}}
\put(   0.023, 172.163){\line(-1,0){3}}
\put(  -4.977, 172.163){\makebox(0,0)[r]{ .20}}
\put(   0.023, 215.157){\line(-1,0){3}}
\put(  -4.977, 215.157){\makebox(0,0)[r]{ .25}}
%\put(   0.023,   0.000){\circle*{1}}
%\put(   0.024,   0.000){\circle*{1}}
%\put(   0.026,   0.000){\circle*{1}}
%\put(   0.027,   0.000){\circle*{1}}
 \put(   0.029,   0.000){\circle*{1}}
 \put(   0.031,   0.000){\circle*{1}}
%\put(   0.033,   0.000){\circle*{1}}
%\put(   0.035,   0.000){\circle*{1}}
%\put(   0.037,   0.000){\circle*{1}}
%\put(   0.040,   0.000){\circle*{1}}
 \put(   0.042,   0.000){\circle*{1}}
 \put(   0.045,   0.000){\circle*{1}}
%\put(   0.048,   0.000){\circle*{1}}
%\put(   0.051,   0.000){\circle*{1}}
%\put(   0.054,   0.000){\circle*{1}}
%\put(   0.057,   0.000){\circle*{1}}
 \put(   0.061,   0.000){\circle*{1}}
 \put(   0.065,   0.000){\circle*{1}}
%\put(   0.069,   0.000){\circle*{1}}
%\put(   0.074,   0.000){\circle*{1}}
%\put(   0.079,   0.000){\circle*{1}}
%\put(   0.084,   0.000){\circle*{1}}
 \put(   0.089,   0.000){\circle*{1}}
 \put(   0.095,   0.000){\circle*{1}}
%\put(   0.101,   0.000){\circle*{1}}
%\put(   0.107,   0.000){\circle*{1}}
%\put(   0.114,   0.000){\circle*{1}}
%\put(   0.122,   0.000){\circle*{1}}
 \put(   0.130,   0.000){\circle*{1}}
 \put(   0.138,   0.000){\circle*{1}}
%\put(   0.147,   0.000){\circle*{1}}
%\put(   0.156,   0.000){\circle*{1}}
%\put(   0.166,   0.000){\circle*{1}}
%\put(   0.177,   0.000){\circle*{1}}
 \put(   0.189,   0.000){\circle*{1}}
 \put(   0.201,   0.000){\circle*{1}}
%\put(   0.214,   0.000){\circle*{1}}
%\put(   0.227,   0.000){\circle*{1}}
%\put(   0.242,   0.000){\circle*{1}}
%\put(   0.258,   0.000){\circle*{1}}
 \put(   0.274,   0.000){\circle*{1}}
 \put(   0.292,   0.000){\circle*{1}}
%\put(   0.311,   0.000){\circle*{1}}
%\put(   0.331,   0.000){\circle*{1}}
%\put(   0.352,   0.000){\circle*{1}}
%\put(   0.375,   0.000){\circle*{1}}
 \put(   0.399,   0.000){\circle*{1}}
 \put(   0.425,   0.000){\circle*{1}}
%\put(   0.452,   0.000){\circle*{1}}
%\put(   0.481,   0.000){\circle*{1}}
%\put(   0.512,   0.000){\circle*{1}}
%\put(   0.546,   0.000){\circle*{1}}
 \put(   0.581,   0.000){\circle*{1}}
 \put(   0.618,   0.000){\circle*{1}}
%\put(   0.658,   0.000){\circle*{1}}
%\put(   0.700,   0.000){\circle*{1}}
%\put(   0.746,   0.000){\circle*{1}}
%\put(   0.794,   0.000){\circle*{1}}
 \put(   0.845,   0.000){\circle*{1}}
 \put(   0.899,   0.000){\circle*{1}}
%\put(   0.957,   0.000){\circle*{1}}
%\put(   1.019,   0.000){\circle*{1}}
%\put(   1.085,   0.000){\circle*{1}}
%\put(   1.155,   0.000){\circle*{1}}
 \put(   1.229,   0.000){\circle*{1}}
 \put(   1.309,   0.000){\circle*{1}}
%\put(   1.393,   0.000){\circle*{1}}
%\put(   1.483,   0.000){\circle*{1}}
%\put(   1.579,   0.000){\circle*{1}}
%\put(   1.680,   0.000){\circle*{1}}
 \put(   1.789,   0.000){\circle*{1}}
 \put(   1.904,   0.000){\circle*{1}}
%\put(   2.027,   0.000){\circle*{1}}
%\put(   2.158,   0.000){\circle*{1}}
%\put(   2.297,   0.000){\circle*{1}}
%\put(   2.445,   0.000){\circle*{1}}
 \put(   2.603,   0.000){\circle*{1}}
 \put(   2.771,   0.000){\circle*{1}}
%\put(   2.949,   0.000){\circle*{1}}
%\put(   3.139,   0.000){\circle*{1}}
%\put(   3.342,   0.000){\circle*{1}}
%\put(   3.557,   0.000){\circle*{1}}
 \put(   3.787,   0.000){\circle*{1}}
 \put(   4.031,   0.000){\circle*{1}}
%\put(   4.291,   0.000){\circle*{1}}
%\put(   4.568,   0.000){\circle*{1}}
%\put(   4.862,   0.000){\circle*{1}}
%\put(   5.176,   0.000){\circle*{1}}
 \put(   5.510,   0.000){\circle*{1}}
 \put(   5.865,   0.000){\circle*{1}}
%\put(   6.243,   0.000){\circle*{1}}
%\put(   6.646,   0.000){\circle*{1}}
%\put(   7.075,   0.000){\circle*{1}}
%\put(   7.531,   0.000){\circle*{1}}
 \put(   8.017,   0.000){\circle*{1}}
 \put(   8.534,   0.000){\circle*{1}}
%\put(   9.084,   0.000){\circle*{1}}
%\put(   9.670,   0.000){\circle*{1}}
%\put(  10.294,   0.001){\circle*{1}}
%\put(  10.958,   0.001){\circle*{1}}
 \put(  11.664,   0.001){\circle*{1}}
 \put(  12.417,   0.001){\circle*{1}}
%\put(  13.217,   0.002){\circle*{1}}
%\put(  14.070,   0.003){\circle*{1}}
%\put(  14.977,   0.004){\circle*{1}}
%\put(  15.943,   0.005){\circle*{1}}
 \put(  16.457,   0.006){\circle*{1}}
 \put(  16.971,   0.006){\circle*{1}}
%\put(  17.519,   0.008){\circle*{1}}
%\put(  18.066,   0.009){\circle*{1}}
%\put(  18.649,   0.010){\circle*{1}}
%\put(  19.231,   0.012){\circle*{1}}
 \put(  19.851,   0.014){\circle*{1}}
 \put(  20.471,   0.016){\circle*{1}}
%\put(  21.132,   0.018){\circle*{1}}
%\put(  21.792,   0.021){\circle*{1}}
%\put(  22.494,   0.024){\circle*{1}}
%\put(  23.197,   0.028){\circle*{1}}
 \put(  23.945,   0.033){\circle*{1}}
 \put(  24.693,   0.037){\circle*{1}}
%\put(  25.490,   0.043){\circle*{1}}
%\put(  26.286,   0.050){\circle*{1}}
%\put(  27.133,   0.058){\circle*{1}}
%\put(  27.981,   0.066){\circle*{1}}
 \put(  28.883,   0.077){\circle*{1}}
 \put(  29.786,   0.087){\circle*{1}}
%\put(  30.746,   0.102){\circle*{1}}
%\put(  31.707,   0.116){\circle*{1}}
%\put(  32.388,   0.128){\circle*{1}}
%\put(  33.070,   0.140){\circle*{1}}
 \put(  33.752,   0.153){\circle*{1}}
 \put(  34.477,   0.169){\circle*{1}}
%\put(  35.203,   0.185){\circle*{1}}
%\put(  35.928,   0.201){\circle*{1}}
%\put(  36.701,   0.222){\circle*{1}}
%\put(  37.473,   0.243){\circle*{1}}
 \put(  38.246,   0.264){\circle*{1}}
 \put(  39.068,   0.292){\circle*{1}}
%\put(  39.890,   0.319){\circle*{1}}
%\put(  40.712,   0.346){\circle*{1}}
%\put(  41.587,   0.382){\circle*{1}}
%\put(  42.463,   0.417){\circle*{1}}
 \put(  43.338,   0.453){\circle*{1}}
 \put(  44.270,   0.498){\circle*{1}}
%\put(  45.201,   0.544){\circle*{1}}
%\put(  46.133,   0.589){\circle*{1}}
%\put(  47.125,   0.648){\circle*{1}}
%\put(  48.117,   0.706){\circle*{1}}
 \put(  49.108,   0.765){\circle*{1}}
 \put(  49.900,   0.821){\circle*{1}}
%\put(  50.692,   0.877){\circle*{1}}
%\put(  51.484,   0.933){\circle*{1}}
%\put(  52.276,   0.989){\circle*{1}}
%\put(  53.118,   1.060){\circle*{1}}
 \put(  53.961,   1.131){\circle*{1}}
 \put(  54.804,   1.202){\circle*{1}}
%\put(  55.647,   1.273){\circle*{1}}
%\put(  56.544,   1.362){\circle*{1}}
%\put(  57.442,   1.452){\circle*{1}}
%\put(  58.339,   1.542){\circle*{1}}
 \put(  59.236,   1.631){\circle*{1}}
 \put(  60.191,   1.744){\circle*{1}}
%\put(  61.146,   1.856){\circle*{1}}
%\put(  62.101,   1.969){\circle*{1}}
%\put(  63.056,   2.081){\circle*{1}}
%\put(  63.870,   2.193){\circle*{1}}
 \put(  64.683,   2.305){\circle*{1}}
 \put(  65.496,   2.417){\circle*{1}}
%\put(  66.310,   2.530){\circle*{1}}
%\put(  67.123,   2.642){\circle*{1}}
%\put(  67.989,   2.780){\circle*{1}}
%\put(  68.855,   2.919){\circle*{1}}
 \put(  69.721,   3.057){\circle*{1}}
 \put(  70.586,   3.196){\circle*{1}}
%\put(  71.452,   3.334){\circle*{1}}
%\put(  72.374,   3.504){\circle*{1}}
%\put(  73.296,   3.674){\circle*{1}}
%\put(  74.217,   3.844){\circle*{1}}
 \put(  75.139,   4.013){\circle*{1}}
 \put(  76.061,   4.183){\circle*{1}}
%\put(  76.878,   4.355){\circle*{1}}
%\put(  77.696,   4.527){\circle*{1}}
%\put(  78.513,   4.699){\circle*{1}}
%\put(  79.331,   4.871){\circle*{1}}
 \put(  80.148,   5.042){\circle*{1}}
 \put(  80.966,   5.214){\circle*{1}}
%\put(  81.836,   5.421){\circle*{1}}
%\put(  82.707,   5.628){\circle*{1}}
%\put(  83.577,   5.834){\circle*{1}}
%\put(  84.447,   6.041){\circle*{1}}
 \put(  85.318,   6.248){\circle*{1}}
 \put(  86.188,   6.454){\circle*{1}}
%\put(  87.114,   6.700){\circle*{1}}
%\put(  88.041,   6.946){\circle*{1}}
%\put(  88.967,   7.192){\circle*{1}}
%\put(  89.894,   7.438){\circle*{1}}
 \put(  90.820,   7.684){\circle*{1}}
 \put(  91.746,   7.929){\circle*{1}}
%\put(  92.592,   8.177){\circle*{1}}
%\put(  93.437,   8.425){\circle*{1}}
%\put(  94.282,   8.672){\circle*{1}}
%\put(  95.128,   8.920){\circle*{1}}
 \put(  95.973,   9.168){\circle*{1}}
 \put(  96.818,   9.415){\circle*{1}}
%\put(  97.664,   9.663){\circle*{1}}
%\put(  98.563,   9.950){\circle*{1}}
%\put(  99.463,  10.237){\circle*{1}}
%\put( 100.363,  10.524){\circle*{1}}
 \put( 101.263,  10.812){\circle*{1}}
 \put( 102.163,  11.099){\circle*{1}}
%\put( 103.063,  11.386){\circle*{1}}
%\put( 103.962,  11.673){\circle*{1}}
%\put( 104.801,  11.961){\circle*{1}}
%\put( 105.639,  12.248){\circle*{1}}
 \put( 106.477,  12.535){\circle*{1}}
 \put( 107.315,  12.823){\circle*{1}}
%\put( 108.153,  13.110){\circle*{1}}
%\put( 108.991,  13.397){\circle*{1}}
%\put( 109.829,  13.685){\circle*{1}}
%\put( 110.667,  13.972){\circle*{1}}
 \put( 111.560,  14.295){\circle*{1}}
 \put( 112.452,  14.619){\circle*{1}}
%\put( 113.344,  14.942){\circle*{1}}
%\put( 114.236,  15.266){\circle*{1}}
%\put( 115.128,  15.589){\circle*{1}}
%\put( 116.020,  15.913){\circle*{1}}
 \put( 116.913,  16.236){\circle*{1}}
 \put( 117.805,  16.560){\circle*{1}}
%\put( 118.649,  16.878){\circle*{1}}
%\put( 119.493,  17.197){\circle*{1}}
%\put( 120.337,  17.515){\circle*{1}}
%\put( 121.182,  17.834){\circle*{1}}
 \put( 122.026,  18.152){\circle*{1}}
 \put( 122.870,  18.471){\circle*{1}}
%\put( 123.714,  18.789){\circle*{1}}
%\put( 124.558,  19.108){\circle*{1}}
%\put( 125.403,  19.426){\circle*{1}}
%\put( 126.301,  19.773){\circle*{1}}
 \put( 127.200,  20.120){\circle*{1}}
 \put( 128.098,  20.467){\circle*{1}}
%\put( 128.997,  20.814){\circle*{1}}
%\put( 129.896,  21.161){\circle*{1}}
%\put( 130.794,  21.508){\circle*{1}}
%\put( 131.693,  21.855){\circle*{1}}
 \put( 132.592,  22.202){\circle*{1}}
 \put( 133.490,  22.549){\circle*{1}}
%\put( 134.351,  22.884){\circle*{1}}
%\put( 135.212,  23.218){\circle*{1}}
%\put( 136.073,  23.553){\circle*{1}}
%\put( 136.934,  23.887){\circle*{1}}
 \put( 137.795,  24.222){\circle*{1}}
 \put( 138.656,  24.557){\circle*{1}}
%\put( 139.517,  24.891){\circle*{1}}
%\put( 140.378,  25.226){\circle*{1}}
%\put( 141.239,  25.560){\circle*{1}}
%\put( 142.100,  25.895){\circle*{1}}
 \put( 143.016,  26.248){\circle*{1}}
 \put( 143.933,  26.602){\circle*{1}}
%\put( 144.849,  26.955){\circle*{1}}
%\put( 145.766,  27.309){\circle*{1}}
%\put( 146.682,  27.662){\circle*{1}}
%\put( 147.599,  28.015){\circle*{1}}
 \put( 148.515,  28.369){\circle*{1}}
 \put( 149.431,  28.722){\circle*{1}}
%\put( 150.348,  29.075){\circle*{1}}
%\put( 151.264,  29.429){\circle*{1}}
%\put( 152.151,  29.765){\circle*{1}}
%\put( 153.038,  30.100){\circle*{1}}
 \put( 153.925,  30.436){\circle*{1}}
 \put( 154.812,  30.772){\circle*{1}}
%\put( 155.699,  31.108){\circle*{1}}
%\put( 156.586,  31.443){\circle*{1}}
%\put( 157.473,  31.779){\circle*{1}}
%\put( 158.359,  32.115){\circle*{1}}
 \put( 159.246,  32.451){\circle*{1}}
 \put( 160.133,  32.786){\circle*{1}}
%\put( 161.020,  33.122){\circle*{1}}
%\put( 161.885,  33.443){\circle*{1}}
%\put( 162.751,  33.764){\circle*{1}}
%\put( 163.616,  34.084){\circle*{1}}
 \put( 164.482,  34.405){\circle*{1}}
 \put( 165.347,  34.726){\circle*{1}}
%\put( 166.213,  35.046){\circle*{1}}
%\put( 167.078,  35.367){\circle*{1}}
%\put( 167.943,  35.688){\circle*{1}}
%\put( 168.809,  36.009){\circle*{1}}
 \put( 169.674,  36.329){\circle*{1}}
 \put( 170.540,  36.650){\circle*{1}}
%\put( 171.405,  36.971){\circle*{1}}
%\put( 172.326,  37.308){\circle*{1}}
%\put( 173.247,  37.645){\circle*{1}}
%\put( 174.169,  37.981){\circle*{1}}
 \put( 175.090,  38.318){\circle*{1}}
 \put( 176.011,  38.655){\circle*{1}}
%\put( 176.932,  38.992){\circle*{1}}
%\put( 177.854,  39.329){\circle*{1}}
%\put( 178.775,  39.666){\circle*{1}}
%\put( 179.696,  40.003){\circle*{1}}
 \put( 180.617,  40.339){\circle*{1}}
 \put( 181.538,  40.676){\circle*{1}}
%\put( 182.460,  41.013){\circle*{1}}
%\put( 183.365,  41.347){\circle*{1}}
%\put( 184.270,  41.680){\circle*{1}}
%\put( 185.175,  42.013){\circle*{1}}
 \put( 186.080,  42.347){\circle*{1}}
 \put( 186.986,  42.680){\circle*{1}}
%\put( 187.891,  43.013){\circle*{1}}
%\put( 188.796,  43.347){\circle*{1}}
%\put( 189.701,  43.680){\circle*{1}}
%\put( 190.606,  44.014){\circle*{1}}
 \put( 191.512,  44.347){\circle*{1}}
 \put( 192.417,  44.680){\circle*{1}}
%\put( 193.322,  45.014){\circle*{1}}
%\put( 194.227,  45.347){\circle*{1}}
%\put( 195.122,  45.689){\circle*{1}}
%\put( 196.017,  46.031){\circle*{1}}
 \put( 196.912,  46.374){\circle*{1}}
 \put( 197.806,  46.716){\circle*{1}}
%\put( 198.701,  47.058){\circle*{1}}
%\put( 199.596,  47.400){\circle*{1}}
%\put( 200.491,  47.742){\circle*{1}}
%\put( 201.385,  48.084){\circle*{1}}
 \put( 202.280,  48.426){\circle*{1}}
 \put( 203.175,  48.769){\circle*{1}}
%\put( 204.070,  49.111){\circle*{1}}
%\put( 204.964,  49.453){\circle*{1}}
%\put( 205.859,  49.795){\circle*{1}}
%\put( 206.754,  50.137){\circle*{1}}
 \put( 207.643,  50.501){\circle*{1}}
 \put( 208.532,  50.866){\circle*{1}}
%\put( 209.421,  51.230){\circle*{1}}
%\put( 210.310,  51.594){\circle*{1}}
%\put( 211.199,  51.959){\circle*{1}}
%\put( 212.088,  52.323){\circle*{1}}
 \put( 212.977,  52.687){\circle*{1}}
 \put( 213.866,  53.051){\circle*{1}}
%\put( 214.755,  53.416){\circle*{1}}
%\put( 215.644,  53.780){\circle*{1}}
%\put( 216.532,  54.144){\circle*{1}}
%\put( 217.421,  54.508){\circle*{1}}
 \put( 218.310,  54.873){\circle*{1}}
 \put( 219.199,  55.237){\circle*{1}}
%\put( 220.088,  55.601){\circle*{1}}
%\put( 220.976,  55.999){\circle*{1}}
%\put( 221.863,  56.397){\circle*{1}}
%\put( 222.750,  56.795){\circle*{1}}
 \put( 223.637,  57.193){\circle*{1}}
 \put( 224.524,  57.591){\circle*{1}}
%\put( 225.411,  57.989){\circle*{1}}
%\put( 226.298,  58.387){\circle*{1}}
%\put( 227.186,  58.785){\circle*{1}}
%\put( 228.073,  59.183){\circle*{1}}
 \put( 228.960,  59.581){\circle*{1}}
 \put( 229.847,  59.979){\circle*{1}}
%\put( 230.734,  60.377){\circle*{1}}
%\put( 231.621,  60.775){\circle*{1}}
%\put( 232.509,  61.173){\circle*{1}}
%\put( 233.396,  61.571){\circle*{1}}
 \put( 234.283,  61.969){\circle*{1}}
 \put( 235.172,  62.407){\circle*{1}}
%\put( 236.060,  62.846){\circle*{1}}
%\put( 236.949,  63.284){\circle*{1}}
%\put( 237.838,  63.723){\circle*{1}}
%\put( 238.727,  64.161){\circle*{1}}
 \put( 239.616,  64.600){\circle*{1}}
 \put( 240.505,  65.038){\circle*{1}}
%\put( 241.393,  65.476){\circle*{1}}
%\put( 242.282,  65.915){\circle*{1}}
%\put( 243.171,  66.353){\circle*{1}}
%\put( 244.060,  66.792){\circle*{1}}
 \put( 244.949,  67.230){\circle*{1}}
 \put( 245.838,  67.669){\circle*{1}}
%\put( 246.726,  68.107){\circle*{1}}
%\put( 247.615,  68.546){\circle*{1}}
%\put( 248.504,  68.984){\circle*{1}}
%\put( 249.393,  69.422){\circle*{1}}
 \put( 250.239,  69.876){\circle*{1}}
 \put( 251.086,  70.330){\circle*{1}}
%\put( 251.932,  70.784){\circle*{1}}
%\put( 252.779,  71.238){\circle*{1}}
%\put( 253.626,  71.692){\circle*{1}}
%\put( 254.472,  72.146){\circle*{1}}
 \put( 255.319,  72.600){\circle*{1}}
 \put( 256.165,  73.054){\circle*{1}}
%\put( 257.012,  73.508){\circle*{1}}
%\put( 257.858,  73.962){\circle*{1}}
%\put( 258.705,  74.416){\circle*{1}}
%\put( 259.551,  74.870){\circle*{1}}
 \put( 260.398,  75.324){\circle*{1}}
 \put( 261.244,  75.778){\circle*{1}}
%\put( 262.091,  76.232){\circle*{1}}
%\put( 262.938,  76.686){\circle*{1}}
%\put( 263.784,  77.140){\circle*{1}}
%\put( 264.631,  77.594){\circle*{1}}
 \put( 265.477,  78.048){\circle*{1}}
 \put( 266.333,  78.539){\circle*{1}}
%\put( 267.189,  79.030){\circle*{1}}
%\put( 268.046,  79.521){\circle*{1}}
%\put( 268.902,  80.011){\circle*{1}}
%\put( 269.758,  80.502){\circle*{1}}
 \put( 270.614,  80.993){\circle*{1}}
 \put( 271.470,  81.484){\circle*{1}}
%\put( 272.326,  81.975){\circle*{1}}
%\put( 273.182,  82.466){\circle*{1}}
%\put( 274.038,  82.956){\circle*{1}}
%\put( 274.894,  83.447){\circle*{1}}
 \put( 275.750,  83.938){\circle*{1}}
 \put( 276.606,  84.429){\circle*{1}}
%\put( 277.463,  84.920){\circle*{1}}
%\put( 278.319,  85.410){\circle*{1}}
\put(   0.023,   0.000){\circle*{1}}
\put(   0.024,   0.000){\circle*{1}}
\put(   0.026,   0.000){\circle*{1}}
\put(   0.027,   0.000){\circle*{1}}
\put(   0.029,   0.000){\circle*{1}}
\put(   0.031,   0.000){\circle*{1}}
\put(   0.033,   0.000){\circle*{1}}
\put(   0.035,   0.000){\circle*{1}}
\put(   0.037,   0.000){\circle*{1}}
\put(   0.040,   0.000){\circle*{1}}
\put(   0.042,   0.000){\circle*{1}}
\put(   0.045,   0.000){\circle*{1}}
\put(   0.048,   0.000){\circle*{1}}
\put(   0.051,   0.000){\circle*{1}}
\put(   0.054,   0.000){\circle*{1}}
\put(   0.057,   0.000){\circle*{1}}
\put(   0.061,   0.000){\circle*{1}}
\put(   0.065,   0.000){\circle*{1}}
\put(   0.069,   0.000){\circle*{1}}
\put(   0.074,   0.000){\circle*{1}}
\put(   0.079,   0.000){\circle*{1}}
\put(   0.084,   0.000){\circle*{1}}
\put(   0.089,   0.000){\circle*{1}}
\put(   0.095,   0.000){\circle*{1}}
\put(   0.101,   0.000){\circle*{1}}
\put(   0.107,   0.000){\circle*{1}}
\put(   0.114,   0.000){\circle*{1}}
\put(   0.122,   0.000){\circle*{1}}
\put(   0.130,   0.000){\circle*{1}}
\put(   0.138,   0.000){\circle*{1}}
\put(   0.147,   0.000){\circle*{1}}
\put(   0.156,   0.000){\circle*{1}}
\put(   0.166,   0.000){\circle*{1}}
\put(   0.177,   0.000){\circle*{1}}
\put(   0.189,   0.000){\circle*{1}}
\put(   0.201,   0.000){\circle*{1}}
\put(   0.214,   0.000){\circle*{1}}
\put(   0.227,   0.000){\circle*{1}}
\put(   0.242,   0.000){\circle*{1}}
\put(   0.258,   0.000){\circle*{1}}
\put(   0.274,   0.000){\circle*{1}}
\put(   0.292,   0.000){\circle*{1}}
\put(   0.311,   0.000){\circle*{1}}
\put(   0.331,   0.000){\circle*{1}}
\put(   0.352,   0.000){\circle*{1}}
\put(   0.375,   0.000){\circle*{1}}
\put(   0.399,   0.000){\circle*{1}}
\put(   0.425,   0.000){\circle*{1}}
\put(   0.452,   0.000){\circle*{1}}
\put(   0.481,   0.000){\circle*{1}}
\put(   0.512,   0.000){\circle*{1}}
\put(   0.546,   0.000){\circle*{1}}
\put(   0.581,   0.000){\circle*{1}}
\put(   0.618,   0.000){\circle*{1}}
\put(   0.658,   0.000){\circle*{1}}
\put(   0.700,   0.000){\circle*{1}}
\put(   0.746,   0.000){\circle*{1}}
\put(   0.794,  -0.001){\circle*{1}}
\put(   0.845,  -0.001){\circle*{1}}
\put(   0.899,  -0.001){\circle*{1}}
\put(   0.957,  -0.001){\circle*{1}}
\put(   1.019,  -0.001){\circle*{1}}
\put(   1.085,  -0.002){\circle*{1}}
\put(   1.155,  -0.002){\circle*{1}}
\put(   1.229,  -0.003){\circle*{1}}
\put(   1.309,  -0.003){\circle*{1}}
\put(   1.393,  -0.004){\circle*{1}}
\put(   1.483,  -0.005){\circle*{1}}
\put(   1.579,  -0.006){\circle*{1}}
\put(   1.680,  -0.008){\circle*{1}}
\put(   1.789,  -0.010){\circle*{1}}
\put(   1.904,  -0.012){\circle*{1}}
\put(   2.027,  -0.015){\circle*{1}}
\put(   2.158,  -0.019){\circle*{1}}
\put(   2.297,  -0.023){\circle*{1}}
\put(   2.445,  -0.029){\circle*{1}}
\put(   2.603,  -0.036){\circle*{1}}
\put(   2.771,  -0.044){\circle*{1}}
\put(   2.949,  -0.054){\circle*{1}}
\put(   3.139,  -0.067){\circle*{1}}
\put(   3.342,  -0.082){\circle*{1}}
\put(   3.557,  -0.101){\circle*{1}}
\put(   3.787,  -0.123){\circle*{1}}
\put(   4.031,  -0.150){\circle*{1}}
\put(   4.291,  -0.183){\circle*{1}}
\put(   4.568,  -0.222){\circle*{1}}
\put(   4.862,  -0.269){\circle*{1}}
\put(   5.176,  -0.325){\circle*{1}}
\put(   5.510,  -0.391){\circle*{1}}
\put(   5.865,  -0.470){\circle*{1}}
\put(   6.243,  -0.562){\circle*{1}}
\put(   6.646,  -0.671){\circle*{1}}
\put(   7.075,  -0.797){\circle*{1}}
\put(   7.531,  -0.943){\circle*{1}}
\put(   8.017,  -1.112){\circle*{1}}
\put(   8.534,  -1.306){\circle*{1}}
\put(   9.084,  -1.526){\circle*{1}}
\put(   9.670,  -1.775){\circle*{1}}
\put(  10.294,  -2.054){\circle*{1}}
\put(  10.958,  -2.365){\circle*{1}}
\put(  11.664,  -2.708){\circle*{1}}
\put(  12.417,  -3.083){\circle*{1}}
\put(  13.217,  -3.489){\circle*{1}}
\put(  14.070,  -3.925){\circle*{1}}
\put(  14.523,  -4.155){\circle*{1}}
\put(  14.977,  -4.386){\circle*{1}}
\put(  15.460,  -4.627){\circle*{1}}
\put(  15.943,  -4.868){\circle*{1}}
\put(  16.457,  -5.116){\circle*{1}}
\put(  16.971,  -5.364){\circle*{1}}
\put(  17.519,  -5.616){\circle*{1}}
\put(  18.066,  -5.867){\circle*{1}}
\put(  18.649,  -6.117){\circle*{1}}
\put(  19.231,  -6.367){\circle*{1}}
\put(  19.851,  -6.609){\circle*{1}}
\put(  20.471,  -6.852){\circle*{1}}
\put(  21.132,  -7.081){\circle*{1}}
\put(  21.792,  -7.310){\circle*{1}}
\put(  22.494,  -7.518){\circle*{1}}
\put(  23.197,  -7.726){\circle*{1}}
\put(  23.945,  -7.906){\circle*{1}}
\put(  24.693,  -8.086){\circle*{1}}
\put(  25.490,  -8.230){\circle*{1}}
\put(  26.286,  -8.374){\circle*{1}}
\put(  27.133,  -8.475){\circle*{1}}
\put(  27.981,  -8.575){\circle*{1}}
\put(  28.883,  -8.625){\circle*{1}}
\put(  29.786,  -8.674){\circle*{1}}
\put(  30.746,  -8.666){\circle*{1}}
\put(  31.707,  -8.658){\circle*{1}}
\put(  32.388,  -8.611){\circle*{1}}
\put(  33.070,  -8.563){\circle*{1}}
\put(  33.752,  -8.515){\circle*{1}}
\put(  34.477,  -8.422){\circle*{1}}
\put(  35.203,  -8.329){\circle*{1}}
\put(  35.928,  -8.236){\circle*{1}}
\put(  36.701,  -8.095){\circle*{1}}
\put(  37.473,  -7.955){\circle*{1}}
\put(  38.246,  -7.814){\circle*{1}}
\put(  39.068,  -7.624){\circle*{1}}
\put(  39.890,  -7.434){\circle*{1}}
\put(  40.712,  -7.244){\circle*{1}}
\put(  41.587,  -7.004){\circle*{1}}
\put(  42.463,  -6.765){\circle*{1}}
\put(  43.338,  -6.525){\circle*{1}}
\put(  44.270,  -6.236){\circle*{1}}
\put(  45.201,  -5.947){\circle*{1}}
\put(  46.133,  -5.658){\circle*{1}}
\put(  46.877,  -5.406){\circle*{1}}
\put(  47.621,  -5.153){\circle*{1}}
\put(  48.365,  -4.900){\circle*{1}}
\put(  49.108,  -4.648){\circle*{1}}
\put(  49.900,  -4.361){\circle*{1}}
\put(  50.692,  -4.074){\circle*{1}}
\put(  51.484,  -3.787){\circle*{1}}
\put(  52.276,  -3.500){\circle*{1}}
\put(  53.118,  -3.180){\circle*{1}}
\put(  53.961,  -2.861){\circle*{1}}
\put(  54.804,  -2.542){\circle*{1}}
\put(  55.647,  -2.223){\circle*{1}}
\put(  56.544,  -1.874){\circle*{1}}
\put(  57.442,  -1.525){\circle*{1}}
\put(  58.339,  -1.176){\circle*{1}}
\put(  59.236,  -0.827){\circle*{1}}
\put(  60.000,  -0.527){\circle*{1}}
\put(  60.764,  -0.226){\circle*{1}}
\put(  61.528,   0.074){\circle*{1}}
\put(  62.292,   0.374){\circle*{1}}
\put(  63.056,   0.675){\circle*{1}}
\put(  63.870,   0.993){\circle*{1}}
\put(  64.683,   1.312){\circle*{1}}
\put(  65.496,   1.631){\circle*{1}}
\put(  66.310,   1.950){\circle*{1}}
\put(  67.123,   2.269){\circle*{1}}
\put(  67.989,   2.604){\circle*{1}}
\put(  68.855,   2.938){\circle*{1}}
\put(  69.721,   3.273){\circle*{1}}
\put(  70.586,   3.608){\circle*{1}}
\put(  71.452,   3.943){\circle*{1}}
\put(  72.374,   4.290){\circle*{1}}
\put(  73.296,   4.638){\circle*{1}}
\put(  74.217,   4.986){\circle*{1}}
\put(  75.139,   5.334){\circle*{1}}
\put(  76.061,   5.682){\circle*{1}}
\put(  76.878,   5.981){\circle*{1}}
\put(  77.696,   6.280){\circle*{1}}
\put(  78.513,   6.579){\circle*{1}}
\put(  79.331,   6.878){\circle*{1}}
\put(  80.148,   7.177){\circle*{1}}
\put(  80.966,   7.476){\circle*{1}}
\put(  81.836,   7.782){\circle*{1}}
\put(  82.707,   8.089){\circle*{1}}
\put(  83.577,   8.395){\circle*{1}}
\put(  84.447,   8.702){\circle*{1}}
\put(  85.318,   9.008){\circle*{1}}
\put(  86.188,   9.315){\circle*{1}}
\put(  87.114,   9.628){\circle*{1}}
\put(  88.041,   9.941){\circle*{1}}
\put(  88.967,  10.254){\circle*{1}}
\put(  89.894,  10.568){\circle*{1}}
\put(  90.820,  10.881){\circle*{1}}
\put(  91.746,  11.194){\circle*{1}}
\put(  92.592,  11.468){\circle*{1}}
\put(  93.437,  11.743){\circle*{1}}
\put(  94.282,  12.017){\circle*{1}}
\put(  95.128,  12.291){\circle*{1}}
\put(  95.973,  12.565){\circle*{1}}
\put(  96.818,  12.840){\circle*{1}}
\put(  97.664,  13.114){\circle*{1}}
\put(  98.563,  13.395){\circle*{1}}
\put(  99.463,  13.676){\circle*{1}}
\put( 100.363,  13.957){\circle*{1}}
\put( 101.263,  14.238){\circle*{1}}
\put( 102.163,  14.519){\circle*{1}}
\put( 103.063,  14.800){\circle*{1}}
\put( 103.962,  15.080){\circle*{1}}
\put( 104.801,  15.334){\circle*{1}}
\put( 105.639,  15.587){\circle*{1}}
\put( 106.477,  15.840){\circle*{1}}
\put( 107.315,  16.094){\circle*{1}}
\put( 108.153,  16.347){\circle*{1}}
\put( 108.991,  16.600){\circle*{1}}
\put( 109.829,  16.853){\circle*{1}}
\put( 110.667,  17.107){\circle*{1}}
\put( 111.560,  17.370){\circle*{1}}
\put( 112.452,  17.633){\circle*{1}}
\put( 113.344,  17.896){\circle*{1}}
\put( 114.236,  18.160){\circle*{1}}
\put( 115.128,  18.423){\circle*{1}}
\put( 116.020,  18.686){\circle*{1}}
\put( 116.913,  18.950){\circle*{1}}
\put( 117.805,  19.213){\circle*{1}}
\put( 118.755,  19.490){\circle*{1}}
\put( 119.704,  19.767){\circle*{1}}
\put( 120.654,  20.044){\circle*{1}}
\put( 121.604,  20.321){\circle*{1}}
\put( 122.553,  20.597){\circle*{1}}
\put( 123.503,  20.874){\circle*{1}}
\put( 124.453,  21.151){\circle*{1}}
\put( 125.403,  21.428){\circle*{1}}
\put( 126.301,  21.690){\circle*{1}}
\put( 127.200,  21.953){\circle*{1}}
\put( 128.098,  22.215){\circle*{1}}
\put( 128.997,  22.477){\circle*{1}}
\put( 129.896,  22.739){\circle*{1}}
\put( 130.794,  23.002){\circle*{1}}
\put( 131.693,  23.264){\circle*{1}}
\put( 132.592,  23.526){\circle*{1}}
\put( 133.490,  23.789){\circle*{1}}
\put( 134.447,  24.072){\circle*{1}}
\put( 135.404,  24.355){\circle*{1}}
\put( 136.360,  24.639){\circle*{1}}
\put( 137.317,  24.922){\circle*{1}}
\put( 138.273,  25.206){\circle*{1}}
\put( 139.230,  25.489){\circle*{1}}
\put( 140.187,  25.773){\circle*{1}}
\put( 141.143,  26.056){\circle*{1}}
\put( 142.100,  26.340){\circle*{1}}
\put( 143.016,  26.619){\circle*{1}}
\put( 143.933,  26.899){\circle*{1}}
\put( 144.849,  27.178){\circle*{1}}
\put( 145.766,  27.458){\circle*{1}}
\put( 146.682,  27.737){\circle*{1}}
\put( 147.599,  28.017){\circle*{1}}
\put( 148.515,  28.296){\circle*{1}}
\put( 149.431,  28.576){\circle*{1}}
\put( 150.348,  28.855){\circle*{1}}
\put( 151.264,  29.135){\circle*{1}}
\put( 152.151,  29.417){\circle*{1}}
\put( 153.038,  29.699){\circle*{1}}
\put( 153.925,  29.980){\circle*{1}}
\put( 154.812,  30.262){\circle*{1}}
\put( 155.699,  30.544){\circle*{1}}
\put( 156.586,  30.826){\circle*{1}}
\put( 157.473,  31.108){\circle*{1}}
\put( 158.359,  31.390){\circle*{1}}
\put( 159.246,  31.672){\circle*{1}}
\put( 160.133,  31.954){\circle*{1}}
\put( 161.020,  32.236){\circle*{1}}
\put( 161.964,  32.552){\circle*{1}}
\put( 162.908,  32.868){\circle*{1}}
\put( 163.852,  33.184){\circle*{1}}
\put( 164.796,  33.500){\circle*{1}}
\put( 165.740,  33.816){\circle*{1}}
\put( 166.685,  34.132){\circle*{1}}
\put( 167.629,  34.448){\circle*{1}}
\put( 168.573,  34.765){\circle*{1}}
\put( 169.517,  35.081){\circle*{1}}
\put( 170.461,  35.397){\circle*{1}}
\put( 171.405,  35.713){\circle*{1}}
\put( 172.326,  36.041){\circle*{1}}
\put( 173.247,  36.368){\circle*{1}}
\put( 174.169,  36.696){\circle*{1}}
\put( 175.090,  37.023){\circle*{1}}
\put( 176.011,  37.351){\circle*{1}}
\put( 176.932,  37.679){\circle*{1}}
\put( 177.854,  38.006){\circle*{1}}
\put( 178.775,  38.334){\circle*{1}}
\put( 179.696,  38.662){\circle*{1}}
\put( 180.617,  38.989){\circle*{1}}
\put( 181.538,  39.317){\circle*{1}}
\put( 182.460,  39.645){\circle*{1}}
\put( 183.365,  39.989){\circle*{1}}
\put( 184.270,  40.333){\circle*{1}}
\put( 185.175,  40.677){\circle*{1}}
\put( 186.080,  41.021){\circle*{1}}
\put( 186.986,  41.365){\circle*{1}}
\put( 187.891,  41.709){\circle*{1}}
\put( 188.796,  42.053){\circle*{1}}
\put( 189.701,  42.397){\circle*{1}}
\put( 190.606,  42.741){\circle*{1}}
\put( 191.512,  43.085){\circle*{1}}
\put( 192.417,  43.430){\circle*{1}}
\put( 193.322,  43.774){\circle*{1}}
\put( 194.227,  44.118){\circle*{1}}
\put( 195.122,  44.482){\circle*{1}}
\put( 196.017,  44.847){\circle*{1}}
\put( 196.912,  45.212){\circle*{1}}
\put( 197.806,  45.577){\circle*{1}}
\put( 198.701,  45.942){\circle*{1}}
\put( 199.596,  46.307){\circle*{1}}
\put( 200.491,  46.671){\circle*{1}}
\put( 201.385,  47.036){\circle*{1}}
\put( 202.280,  47.401){\circle*{1}}
\put( 203.175,  47.766){\circle*{1}}
\put( 204.070,  48.131){\circle*{1}}
\put( 204.964,  48.496){\circle*{1}}
\put( 205.859,  48.860){\circle*{1}}
\put( 206.754,  49.225){\circle*{1}}
\put( 207.643,  49.615){\circle*{1}}
\put( 208.532,  50.004){\circle*{1}}
\put( 209.421,  50.394){\circle*{1}}
\put( 210.310,  50.783){\circle*{1}}
\put( 211.199,  51.172){\circle*{1}}
\put( 212.088,  51.562){\circle*{1}}
\put( 212.977,  51.951){\circle*{1}}
\put( 213.866,  52.341){\circle*{1}}
\put( 214.755,  52.730){\circle*{1}}
\put( 215.644,  53.120){\circle*{1}}
\put( 216.532,  53.509){\circle*{1}}
\put( 217.421,  53.899){\circle*{1}}
\put( 218.310,  54.288){\circle*{1}}
\put( 219.199,  54.678){\circle*{1}}
\put( 220.088,  55.067){\circle*{1}}
\put( 220.976,  55.485){\circle*{1}}
\put( 221.863,  55.902){\circle*{1}}
\put( 222.750,  56.320){\circle*{1}}
\put( 223.637,  56.737){\circle*{1}}
\put( 224.524,  57.155){\circle*{1}}
\put( 225.411,  57.572){\circle*{1}}
\put( 226.298,  57.990){\circle*{1}}
\put( 227.186,  58.407){\circle*{1}}
\put( 228.073,  58.825){\circle*{1}}
\put( 228.960,  59.242){\circle*{1}}
\put( 229.847,  59.660){\circle*{1}}
\put( 230.734,  60.077){\circle*{1}}
\put( 231.621,  60.495){\circle*{1}}
\put( 232.509,  60.912){\circle*{1}}
\put( 233.396,  61.330){\circle*{1}}
\put( 234.283,  61.747){\circle*{1}}
\put( 235.172,  62.196){\circle*{1}}
\put( 236.060,  62.644){\circle*{1}}
\put( 236.949,  63.093){\circle*{1}}
\put( 237.838,  63.541){\circle*{1}}
\put( 238.727,  63.990){\circle*{1}}
\put( 239.616,  64.439){\circle*{1}}
\put( 240.505,  64.887){\circle*{1}}
\put( 241.393,  65.336){\circle*{1}}
\put( 242.282,  65.784){\circle*{1}}
\put( 243.171,  66.233){\circle*{1}}
\put( 244.060,  66.681){\circle*{1}}
\put( 244.949,  67.130){\circle*{1}}
\put( 245.838,  67.578){\circle*{1}}
\put( 246.726,  68.027){\circle*{1}}
\put( 247.615,  68.475){\circle*{1}}
\put( 248.504,  68.924){\circle*{1}}
\put( 249.393,  69.372){\circle*{1}}
\put( 250.239,  69.829){\circle*{1}}
\put( 251.086,  70.285){\circle*{1}}
\put( 251.932,  70.742){\circle*{1}}
\put( 252.779,  71.198){\circle*{1}}
\put( 253.626,  71.655){\circle*{1}}
\put( 254.472,  72.111){\circle*{1}}
\put( 255.319,  72.568){\circle*{1}}
\put( 256.165,  73.024){\circle*{1}}
\put( 257.012,  73.481){\circle*{1}}
\put( 257.858,  73.937){\circle*{1}}
\put( 258.705,  74.394){\circle*{1}}
\put( 259.551,  74.850){\circle*{1}}
\put( 260.398,  75.307){\circle*{1}}
\put( 261.244,  75.763){\circle*{1}}
\put( 262.091,  76.220){\circle*{1}}
\put( 262.938,  76.676){\circle*{1}}
\put( 263.784,  77.133){\circle*{1}}
\put( 264.631,  77.589){\circle*{1}}
\put( 265.477,  78.046){\circle*{1}}
\put( 266.333,  78.537){\circle*{1}}
\put( 267.189,  79.027){\circle*{1}}
\put( 268.046,  79.518){\circle*{1}}
\put( 268.902,  80.009){\circle*{1}}
\put( 269.758,  80.500){\circle*{1}}
\put( 270.614,  80.991){\circle*{1}}
\put( 271.470,  81.482){\circle*{1}}
\put( 272.326,  81.973){\circle*{1}}
\put( 273.182,  82.464){\circle*{1}}
\put( 274.038,  82.955){\circle*{1}}
\put( 274.894,  83.446){\circle*{1}}
\put( 275.750,  83.937){\circle*{1}}
\put( 276.606,  84.428){\circle*{1}}
\put( 277.463,  84.919){\circle*{1}}
\put( 278.319,  85.410){\circle*{1}}
\put( 279.175,  85.901){\circle*{1}}
\put( 280.031,  86.392){\circle*{1}}
\put( 280.887,  86.883){\circle*{1}}
\put( 281.743,  87.374){\circle*{1}}
\put( 282.599,  87.865){\circle*{1}}
\put( 283.428,  88.367){\circle*{1}}
\put( 284.256,  88.869){\circle*{1}}
\put( 285.084,  89.371){\circle*{1}}
\put( 285.913,  89.873){\circle*{1}}
\put( 286.741,  90.376){\circle*{1}}
\put( 287.570,  90.878){\circle*{1}}
\put( 288.398,  91.380){\circle*{1}}
\put( 289.227,  91.882){\circle*{1}}
\put( 290.055,  92.384){\circle*{1}}
\put( 290.884,  92.887){\circle*{1}}
\put( 291.712,  93.389){\circle*{1}}
\put( 292.541,  93.891){\circle*{1}}
\put( 293.369,  94.393){\circle*{1}}
\put( 294.197,  94.895){\circle*{1}}
\put( 295.026,  95.398){\circle*{1}}
\put( 295.854,  95.900){\circle*{1}}
\put( 296.683,  96.402){\circle*{1}}
\put( 297.511,  96.904){\circle*{1}}
\put( 298.340,  97.406){\circle*{1}}
\put( 299.168,  97.909){\circle*{1}}
\put( 299.997,  98.411){\circle*{1}}
\put( 300.825,  98.913){\circle*{1}}
\put( 301.669,  99.450){\circle*{1}}
\put( 302.512,  99.986){\circle*{1}}
\put( 303.356, 100.523){\circle*{1}}
\put( 304.199, 101.059){\circle*{1}}
\put( 305.043, 101.596){\circle*{1}}
\put( 305.886, 102.132){\circle*{1}}
\put( 306.730, 102.669){\circle*{1}}
\put( 307.573, 103.205){\circle*{1}}
\put( 308.417, 103.742){\circle*{1}}
\put( 309.261, 104.278){\circle*{1}}
\put( 310.104, 104.815){\circle*{1}}
\put( 310.948, 105.351){\circle*{1}}
\put( 311.791, 105.888){\circle*{1}}
\put( 312.635, 106.424){\circle*{1}}
\put( 313.478, 106.961){\circle*{1}}
\put( 314.322, 107.497){\circle*{1}}
\put( 315.165, 108.034){\circle*{1}}
\put( 316.009, 108.570){\circle*{1}}
\put( 316.852, 109.107){\circle*{1}}
\put( 317.696, 109.643){\circle*{1}}
\put( 318.540, 110.180){\circle*{1}}
\put( 319.383, 110.716){\circle*{1}}
\put( 320.227, 111.253){\circle*{1}}
\put( 321.053, 111.799){\circle*{1}}
\put( 321.879, 112.346){\circle*{1}}
\put( 322.705, 112.892){\circle*{1}}
\put( 323.531, 113.439){\circle*{1}}
\put( 324.357, 113.985){\circle*{1}}
\put( 325.183, 114.532){\circle*{1}}
\put( 326.009, 115.078){\circle*{1}}
\put( 326.836, 115.625){\circle*{1}}
\put( 327.662, 116.171){\circle*{1}}
\put( 328.488, 116.718){\circle*{1}}
\put( 329.314, 117.264){\circle*{1}}
\put( 330.140, 117.811){\circle*{1}}
\put( 330.966, 118.357){\circle*{1}}
\put( 331.792, 118.904){\circle*{1}}
\put( 332.618, 119.450){\circle*{1}}
\put( 333.444, 119.997){\circle*{1}}
\put( 334.271, 120.543){\circle*{1}}
\put( 335.097, 121.090){\circle*{1}}
\put( 335.923, 121.636){\circle*{1}}
\put( 336.749, 122.183){\circle*{1}}
\put( 337.575, 122.729){\circle*{1}}
\put( 338.401, 123.276){\circle*{1}}
\put( 339.227, 123.822){\circle*{1}}
\put( 340.053, 124.369){\circle*{1}}
\put( 340.880, 124.915){\circle*{1}}
\put( 341.694, 125.470){\circle*{1}}
\put( 342.508, 126.024){\circle*{1}}
\put( 343.322, 126.579){\circle*{1}}
\put( 344.137, 127.133){\circle*{1}}
\put( 344.951, 127.688){\circle*{1}}
\put( 345.765, 128.242){\circle*{1}}
\put( 346.579, 128.797){\circle*{1}}
\put( 347.394, 129.352){\circle*{1}}
\put( 348.208, 129.906){\circle*{1}}
\put( 349.022, 130.461){\circle*{1}}
\put( 349.836, 131.015){\circle*{1}}
\put( 350.651, 131.570){\circle*{1}}
\put( 351.465, 132.124){\circle*{1}}
\put( 352.279, 132.679){\circle*{1}}
\put( 353.093, 133.233){\circle*{1}}
\put( 353.908, 133.788){\circle*{1}}
\put( 354.722, 134.343){\circle*{1}}
\put( 355.536, 134.897){\circle*{1}}
\put( 356.350, 135.452){\circle*{1}}
\put( 357.165, 136.006){\circle*{1}}
\put( 357.979, 136.561){\circle*{1}}
\put( 358.793, 137.115){\circle*{1}}
\put( 359.607, 137.670){\circle*{1}}
\put( 360.422, 138.224){\circle*{1}}
\put( 361.236, 138.779){\circle*{1}}
\put( 362.050, 139.334){\circle*{1}}
\put( 362.864, 139.888){\circle*{1}}
\put( 363.671, 140.447){\circle*{1}}
\put( 364.478, 141.007){\circle*{1}}
\put( 365.285, 141.566){\circle*{1}}
\put( 366.092, 142.125){\circle*{1}}
\put( 366.899, 142.684){\circle*{1}}
\put( 367.706, 143.244){\circle*{1}}
\put( 368.513, 143.803){\circle*{1}}
\put( 369.320, 144.362){\circle*{1}}
\put( 370.127, 144.921){\circle*{1}}
\put( 370.934, 145.481){\circle*{1}}
\put( 371.741, 146.040){\circle*{1}}
\put( 372.548, 146.599){\circle*{1}}
\put( 373.355, 147.158){\circle*{1}}
\put( 374.162, 147.717){\circle*{1}}
\put( 374.969, 148.277){\circle*{1}}
\put( 375.776, 148.836){\circle*{1}}
\put( 376.583, 149.395){\circle*{1}}
\put( 377.390, 149.954){\circle*{1}}
\put( 378.197, 150.514){\circle*{1}}
\put( 379.004, 151.073){\circle*{1}}
\put( 379.811, 151.632){\circle*{1}}
\put( 380.618, 152.191){\circle*{1}}
\put( 381.425, 152.751){\circle*{1}}
\put( 382.232, 153.310){\circle*{1}}
\put( 383.039, 153.869){\circle*{1}}
\put( 383.846, 154.428){\circle*{1}}
\put( 384.653, 154.988){\circle*{1}}
\put( 385.460, 155.547){\circle*{1}}
\end{picture}
\vspace{6cm}
\caption{The radial parts of the $5f_{5/2}$ pseudospinor (dashed line) and
         the large component of the $5f_{5/2}$ spinor (solid line) for the Pb atom
         from work~\protect\cite{Tupitsyn}.}
\label{Pb_5f}
\end{center}
\end{figure}
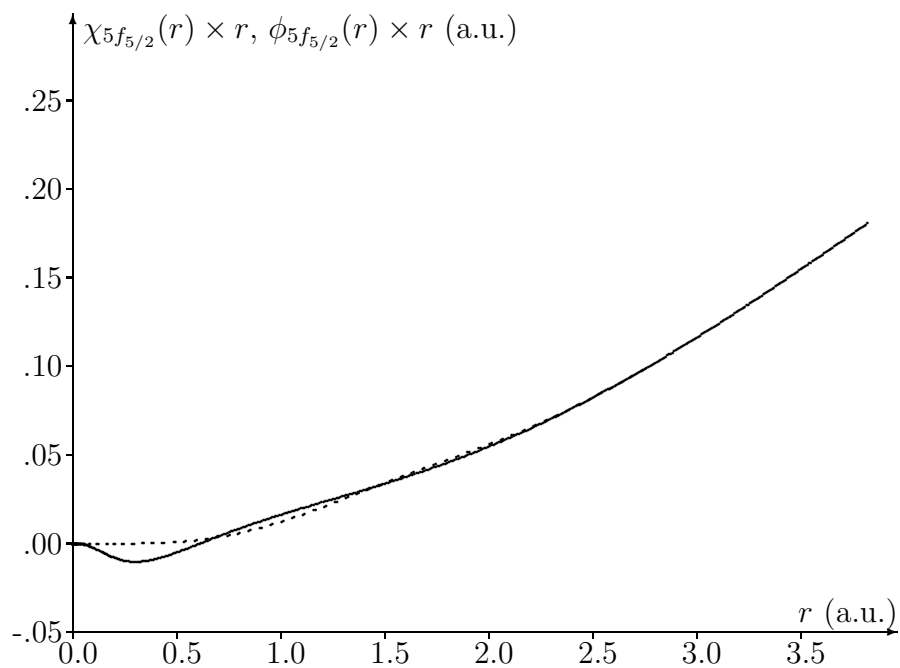

\end{document}